\documentstyle[twocolumn,aps,epsf]{revtex}
\begin{document}
\draft
\twocolumn[\hsize\textwidth\columnwidth\hsize\csname @twocolumnfalse\endcsname
%
%
%

\title{ Low-Energy Properties of Antiferromagnetic
Spin-1/2 Heisenberg Ladders with an Odd Number of Legs}

\author{ Beat Frischmuth, Stephan Haas, German Sierra, and
T.M. Rice}

\address{Institute of Theoretical Physics, ETH H\"onggerberg, CH-8093 Z\"urich,
Switzerland}

\date{\today}
\maketitle

\begin{abstract}
 An effective low-energy description for multi-leg spin-1/2 Heisenberg ladders
with an odd number of legs is proposed. 
Using a newly developed Monte Carlo loop algorithm and exact diagonalization
techniques, the uniform and staggered
magnetic susceptibility and the entropy are calculated for ladders
with 1, 3, and 5 legs.
These systems show a low-temperature scaling behavior similar 
to spin-1/2 chains with longer ranged unfrustrated exchange interactions.
The spinon velocity does not change as the number of legs increases,
but the energy scale parameter decreases markedly.

\end{abstract}

\vskip2pc]
\narrowtext

%
%
Recently, antiferromagnetic Heisenberg spin-$1/2$ ladder attracted 
much interest, following the discovery of a spin gap
in the 2-leg-ladder \cite{dagottoriera}. Also the crossover from the 
single chain to the
two-dimensional (2D) square lattice, obtained by
assembling chains to form ``ladders'' of increasing width, is far from
smooth \cite{dagotto}. Heisenberg ladders with an even number of legs, 
$\rm n_{l}$, have a spin gap and short range correlations, while 
odd-leg ladders have no gap and power-law correlations.  These 
theoretical predictions have been
verified experimentally, in materials such as $(VO)_{2}P_{2}O_{7}$
\cite{johnston} and the homologous series of cuprates
$Sr_{n-1}Cu_{n+1}O_{2n}$ \cite{hiroi}, which contain weakly coupled
arrays of ladders.

Here, we concentrate on odd-leg ladders. Our goal is to derive a
low-energy description in terms of S=1/2 chains with longer range
effective interactions, and examine the evolution with increasing
number of legs, n${\rm _l}$. The
Heisenberg Hamiltonian for ladders is
\begin{equation}
{\rm\label{Js}
H=J_{\parallel} \sum_{\leftrightarrow} \vec{S}_{i,\tau} \vec{S}_{j,\tau}
    + J_{\perp}\sum_{\updownarrow} \vec{S}_{i,\tau} \vec{S}_{i,\tau '},
}
\end{equation}
where i and j enumerate the rungs, $\rm \tau$, $\rm \tau '$ 
label the legs, and the sum marked by 
$\leftrightarrow$ ($\updownarrow$) runs over 
nearest neighbors along legs (rungs). Periodic boundary conditions  
are chosen
along the leg direction and open boundary conditions perpendicular to it. 
For the known materials we expect the superexchange to be 
roughly isotropic (${\rm J_{\perp} = J_{\parallel}}$). However, it is 
educational first to consider the strongly anisotropic limit 
(${\rm J_{\perp} \gg J_{\parallel}}$).

In the completely anisotropic
limit (${\rm J_{\parallel}/J_{\perp}=0}$), each eigenfunction is a direct 
product of one-rung states whose lowest-lying multiplet is a spin doublet,
separated by a gap of order ${\rm J_{\perp}}$ from the first excited state. 
The ground state of the whole system is therefore $\rm 2^{L}$-fold degenerate. 
A finite value of ${\rm J_{\parallel}}$ lifts this degeneracy. 
Our goal is to formulate an effective Hamiltonian, ${\rm H_{eff} }$, in this
$\rm 2^{L}$-dimensional subspace $\cal M$  of rung doublets
which describes the low-energy properties.                
For the case of the 3-leg-ladder, to third order 
in ${\rm J_{\parallel}/J_{\perp}}$, we get
\begin{eqnarray}\label{heff}
\rm
H^{(3)}_{eff} &=& \rm\sum_{j=1}^L[ \sum_{n=1}^3 J_{n} \vec{S}^{tot}_j 
\vec{S}^{tot}_{j + n} + \tilde J ((\vec{S}^{tot}_j\vec{S}^{tot}_{j+3})
(\vec{S}^{tot}_{j+1}\vec{S}^{tot}_{j+2}) \nonumber \\
& & \ \ \ \ \ \ \ \ \rm 
-(\vec{S}^{tot}_j\vec{S}^{tot}_{j+2})(\vec{S}^{tot}_{j+1}\vec{S}^{tot}_{j+3}))]
\end{eqnarray}
where 
${\rm \vec{S}_{j}^{\:\mbox{\scriptsize tot}}=
\vec{S}_{j,1}+\vec{S}_{j,2}+\vec{S}_{j,3}}$ is the total spin of the 
$\rm j^{th}$ rung,  ${\rm J_n = J_{\perp}
\sum_{\lambda} a_{n,\lambda} (\frac{J_{\parallel}}{J_{\perp}})^{\lambda} }$,
with ${\rm a_{1,1}}$=1, ${\rm  a_{1,2}}$= -1/9,
${\rm a_{1,3}}$= -103/243, ${\rm 
a_{2,1}}$=0, ${\rm  a_{2,2}}$=-8/27, ${\rm  a_{2,3}}$=-49/162,
${\rm  a_{3,1}}$=${\rm  a_{3,2}}$=0, ${\rm  a_{3,3}}$=32/243, and ${\rm 
\tilde J=(16/81) J_{\parallel}^3/J_{\perp}^2 }$.

The last term in Eq. (\ref{heff}) will be neglected, since the corrections 
in the energy of the low-lying energy states, due to this term are small.
${\rm H^{(3)}_{eff} }$ has then the form of a single chain with effective 
nearest neighbor (n.n.) coupling ${\rm J_{1}}$, next-nearest neighbor 
(n.n.n.) coupling ${\rm J_{2}}$, and with exchange coupling ${\rm J_{3}}$ 
between rung spins separated by three unit cells.
Therefore the low-lying energy states of the 3-leg-ladder can be 
mapped onto those of a ${\rm J_1-J_2-J_3}$ chain. 
In this effective system the n.n.n. interaction are 
F ($\rm J_2<0$), while the n.n.n.n. interactions are 
AF ($\rm J_3>0$). Consequently, both the second and the 
third term in Eq.(\ref{heff}) enhance the overall AF
quasi-long range order. Note, that the third order corrections in 
Eq.(\ref{heff}) affect ${\rm J_{1}}$ and ${\rm J_{2}}$ strongly since 
the corresponding coefficients are large. So one must perform the 
calculations at least up to third order.

To test ${\rm H^{(3)}_{eff}}$,
we calculate the temperature dependent uniform
susceptibility, ${\rm \chi(T)}$, for 3-leg ladders \cite{ich}
using the Quantum Monte Carlo (QMC) loop algorithm 
$\cite{evertz1,evertz2,wiese}$, and compare to susceptibilities 
obtained for the effective ${\rm J_1-J_2-J_3}$ model. 
We consider large enough systems, such that finite size effects are 
negligible. All results are extrapolated to a Trotter time 
interval $\Delta\tau\rightarrow 0$. Further, we compare 
with recent results obtained by Greven et al. \cite{greven} using the 
same algorithm.

\begin{figure}[h]
\epsfxsize=80mm
\vspace*{-0.5cm}
\epsffile{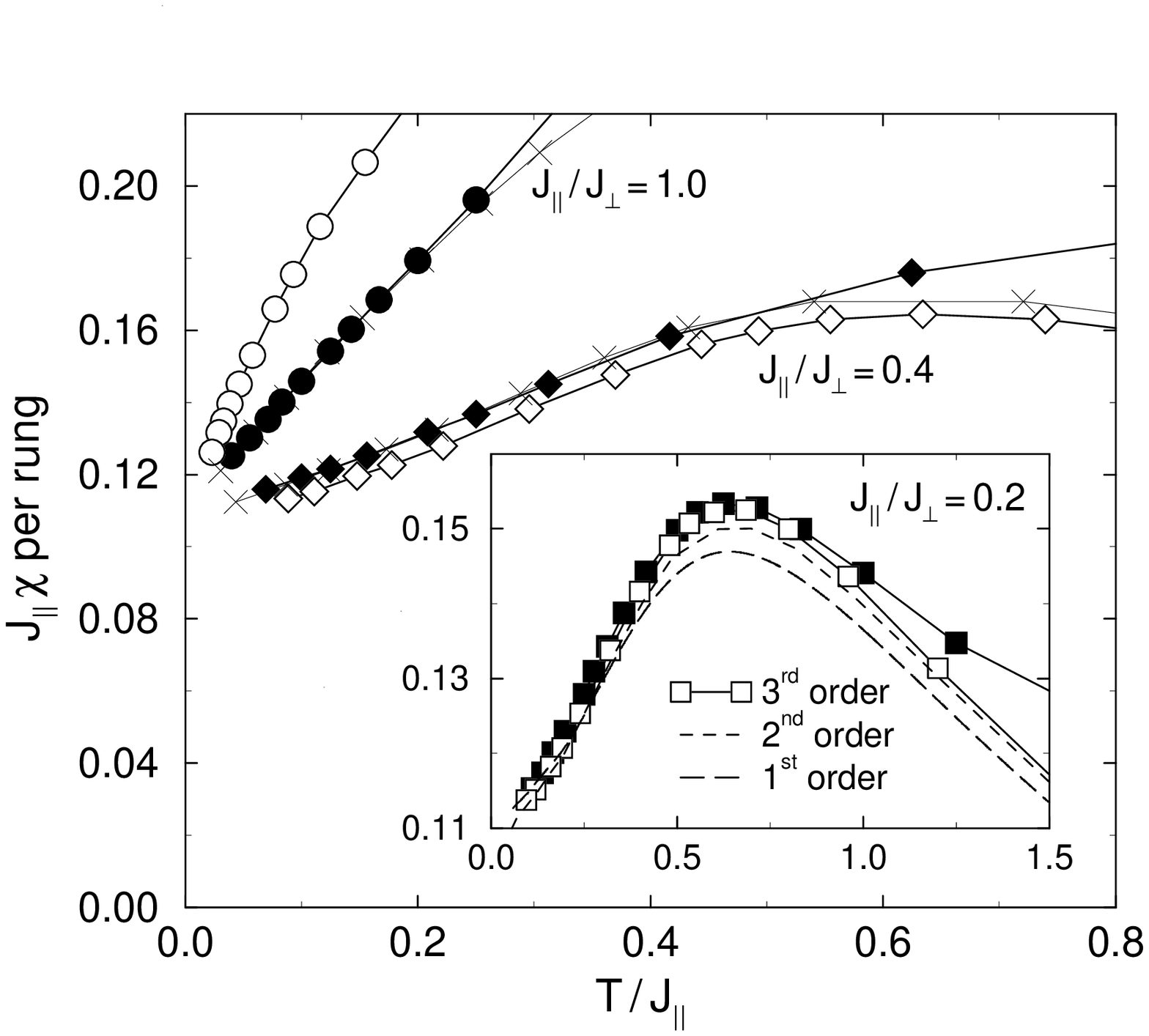}
\caption[*]{Susceptibility of the 3-leg ladder for
different ${\rm J_{\parallel}/J_{\perp}}$ and of the corresponding
effective models. The filled symbols show the data for the 3-leg ladders
and the open symbols those of the corresponding third order effective 
model. The crosses show the susceptibility of the corresponding 
$\rm J_{1}-J_{2}$ chains in the mapping of the 3-leg ladders to 
$\rm J_{1}-J_{2}$ chains (for details see text). The inset shows 
the susceptibility per 
rung of the 3-leg ladder with ${\rm J_{\parallel}/J_{\perp}=0.2}$ 
together with those of the corresponding effective models in first, second 
and third order in ${\rm J_{\parallel}/J_{\perp}}$. The error bars are 
smaller or of the order of the symbols.\label{Fig_susc}}
\end{figure}

At low temperatures, where only the states in $\cal M$ 
are relevant, the susceptibilities of the 3-leg ladders with 
small ${\rm J_{\parallel}/J_{\perp}}$ coincide with those of the 
corresponding ${\rm J_1-J_2-J_3}$ chain. 
This can be seen in the inset of
Fig.\ref{Fig_susc}, where we show the susceptibility per 
rung of the 3-leg ladder with ${\rm J_{\parallel}/J_{\perp}=0.2}$ 
together with the 
susceptibilities of the corresponding effective models in first, second and 
third order in ${\rm J_{\parallel}/J_{\perp}}$. 
While the first order effective 
model (a single chain with only n.n. interactions) gives only a 
qualitative description of the 3-leg ladder at low temperatures, 
${\rm \chi(T)}$ 
of the effective model in third order in ${\rm J_{\parallel}/J_{\perp}}$ 
coincides with the susceptibility per rung of the 3-leg ladder up to a 
crossover temperature. Above this temperature, ${\rm \chi(T)}$ 
of the 3-leg ladder 
is larger, due to the presence of additional states in the 3-leg ladder 
which are not included in the subspace $\cal M$.

\begin{figure}[h]
\epsfxsize=75mm
\vspace*{-0.5cm}
\epsffile{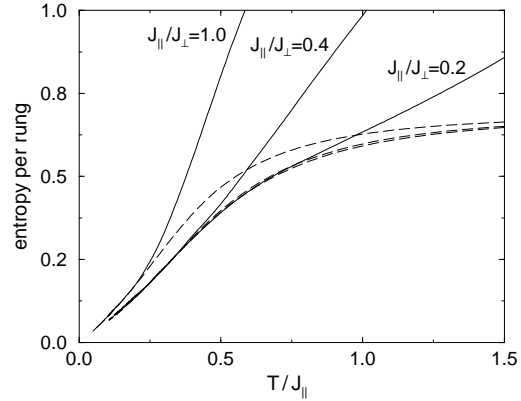}
\caption[*]{Entropy of the 3-leg ladder for 
different $\rm J_{\parallel}/J_{\perp}$ (solid lines) and of the corresponding
effective models (dashed lines).\label{Fig_entropy}}
\end{figure}

The crossover temperature is of the order of the gap $\rm\Delta$ to the 
higher lying states in the 3-leg ladder. It can be estimated best by 
considering the entropy of both the 3-leg ladder and the corresponding 
effective model (Fig. \ref{Fig_entropy}). Just above the crossover 
temperature, the additional states 
lead to a rise ${\rm \propto e^{-\Delta/T}}$ in the entropy. 
Consequently, 
fitting the form ${\rm e^{-\Delta/T}}$ 
to the difference of the entropy per rung 
of the 3-leg ladder and the entropy of the corresponding effective model 
gives an rough estimate of $\rm\Delta$. We find 
${\rm \Delta\approx J_{\perp}}$ more or 
less independent of ${\rm J_{\parallel}/J_{\perp}}$.

From Fig.\ref{Fig_susc},
it is clear that with increasing ${\rm J_{\parallel}/J_{\perp}}$ the 
quality of the description of the 3-leg ladder by the third order 
effective model becomes worse.
The $qualitative$ features of the temperature dependence of the 
susceptibility, however, are correctly given even in the 
isotropic case ${\rm J_{\parallel}/J_{\perp}=1}$, e.g. the slope of 
${\rm \chi(T)}$ 
is increasing, while the zero-temperature value $\rm \chi(0)$ remains 
more or less constant, as ${\rm J_{\parallel}/J_{\perp}}$ increases 
(see Fig.\ref{Fig_susc}). 

Calculating the effective Hamiltonian of the 3-leg ladders to 
${\rm n^{th}}$ order leads to a spin-chain Hamiltonian with interaction 
terms between spins which are separated by up to n unit cells. 
All these interactions are invariant under translations and rotations. 
Consequently, for very low temperatures the 3-leg ladders can be mapped 
onto the $\rm k=1$ Wess-Zumino-Witten (WZW) nonlinear $\sigma$-model 
\cite{ich,eggert}, which is determined by a spinon velocity, $\rm v$, and a 
energy scale parameter, $\rm T_{0}$. For $\rm T \ll T_{0}$, $\rm \chi(T)$
 of this model reads \cite{eggert,nomura} up to: 
$\rm O\left((\ln\: T)^{-3}\right)$
\begin{eqnarray}
\label{1susc}\rm 
\chi(T)=\frac{1}{2\pi v} + \frac{1}{4 \pi v}\left[
\frac{1}{\ln(T_{0}/T)}-\frac{\ln(\ln(T_{0}/T)+1/2)}
{2\ln^{2}(T_{0}/T)}\right]            
\end{eqnarray}
$\rm \chi(T)$ approaches its $\rm T=0$ value $\rm\chi(0)=1/(2\pi v)$ 
with infinite slope. $\rm T_{0}$ should be $\lesssim$
$\rm\Delta$, and characterizes the interactions between the spinons. 
The smaller $\rm T_{0}$, the stronger the interactions, and the faster  
$\rm \chi(T)$ increases with temperature.

The low temperature regime of the universality class of 
spin chains with a rotationally and translationally invariant 
Hamiltonian (to which also the 3-leg ladders belong) is determined by only 
two parameters, v and $\rm T_{0}$. However, the determination of v 
and $\rm T_{0}$ for the 3-leg ladder is difficult. QMC 
calculations cannot be performed down to low enough temperatures such that a 
fit to the above form (\ref{1susc}) gives reliable estimates for v 
and $\rm T_{0}$. Especially $\rm T_{0}$ is considerably underestimated in 
all cases. 

To overcome this problems we first map the 3-leg ladder 
to a ${\rm J_{1}-J_{2}}$ chain and then study the one-to-one mapping of 
this
${\rm J_{1}-J_{2}}$ chain to the WZW model: $\rm (J_{1},\: J_{2})
\leftrightarrow (v,\: T_{0})$. The mapping to ${\rm J_{1}-J_{2}}$ chains
is always possible since the low-T range 
is characterized by only two parameters which can be chosen 
as ${\rm J_{1}}$, ${\rm J_{2}}$ instead of v, $\rm T_{0}$. 
The mapping of the 3-leg ladder to ${\rm J_{1}-J_{2}}$ chains is done as 
follows. For small $\rm J_{\parallel}/J_{\perp}$ we can use the third order 
result for $\rm J_{1}$ and $\rm J_{2}$ (Eq. (\ref{heff})), 
neglecting $\rm J_{3}$, since 
$\rm J_{3}$ is small. Otherwise we fit $\rm \chi(T)$ of 
${\rm J_{1}-J_{2}}$ chains for low T to $\rm \chi(T)$ of the 
3-leg ladder (see Fig. \ref{Fig_susc}) which gives estimates of 
the values $\rm J_{1}$ and $\rm J_{2}$ (see Table \ref{table1}).
\begin{figure}[h]
\epsfxsize=80mm
\vspace*{-0.5cm}
\epsffile{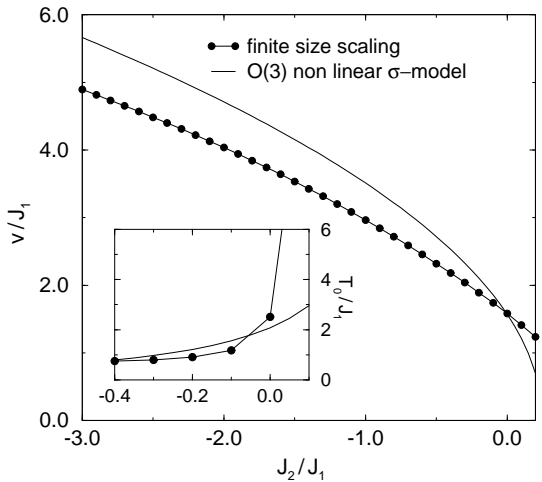}
\caption[*]{Spinon velocity, v, and energy scale parameter, $\rm T_{0}$, 
of the ${\rm J_{1}-J_{2}}$ chains, determined by finite size scaling 
analysis (circles) respectively by mapping to the O(3)-non linear 
$\sigma$-model.\label{Fig_sigma}}
\end{figure}

The mapping $({\rm J_{1}},\; {\rm J_{2}})\leftrightarrow (v,\; T_{0})$ 
can be studied, 
using exact diagonalisation methods. v and $\rm T_{0}$ are determined by 
the finite size scaling of the energy gap between the excited 
state $\rm E(k=\pi,\: S_{z}=1)$ and the ground state 
$\rm E(k=0,\: S_{z}=0)$ 
\cite{nomura}:
\begin{eqnarray}
\label{finitesize}\rm
\lefteqn{\rm E(k=\pi,\: S_{z}=1)-E(k=0,\: S_{z}=0)=}\nonumber \\
& &\rm \frac{\pi v}{L}\left(1-\frac{1}{2 \ln(L/L_{0})}+
\frac{\ln(\ln(L/L_{0})+1/2)}{4 \ln^{2}(L/L_{0})}\right),
\end{eqnarray}
where $\rm E(k,\:S_{z})$ is the lowest energy with wavevector k  
and z-component of
spin $\rm S_{z}$ for a chain of length L. $\rm L_{0}$ is the characteristic
scaling length 
of the chain.
As a consequence of the equivalence of the imaginary time direction 
and the space direction, $\rm L_{0}=v/T_{0}$. Fitting 
Eq.(\ref{finitesize}) for different lengths using exact diagonalisation 
gives v as well as $\rm T_{0}$. The results for different fractions 
${\rm J_{2}/J_{1}}$ are plotted in Fig.\ref{Fig_sigma}.
An alternate possibility to determine v 
\begin{table}
\begin{center}
\begin{tabular}{c | c c | c l}
${\rm J_{\parallel}/J_{\perp}}$ & ${\rm J_{1}/J_{\parallel}}$ & ${\rm J_{2}/J_{\parallel}}$ & $v/{\rm J_{\parallel}}$ & $\rm T_{0}/{\rm J_{\parallel}}$ \\
\hline
0 & 1 & 0 & $\pi/2$ & 2.6 \\
0.1 & 0.985 & -0.033 & 1.61 & 1.64 \\
0.2 & 0.961 & -0.071 & 1.63 & 1.28 \\
0.4 & 0.86 & -0.17 & 1.63 & 0.78 \\
0.6 & 0.76 & -0.30 & 1.65 & 0.57 \\
0.8 & 0.67 & -0.47 & 1.73 & 0.47 \\
1.0 & 0.61 & -0.61 & 1.81 & 0.41 [0.34]\\
\end{tabular}
\end{center}
\caption[*]{For low temperatures, 3-leg ladders can be 
mapped onto ${\rm J_{1}-J_{2}}$ chains. 
The corresponding coupling constants, ${\rm J_{1}}$ 
the 3-leg ladders are also listed. The value enclosed in bracket 
was obtained by fitting the $\rm \xi$-data \cite{greven} to Eq. (\ref{xsi}).
\label{table1}}
\end{table}
\hspace{-0.45cm} and $\rm T_{0}$ is to map
the ${\rm J_{1}-J_{2}}$ chains onto the O(3) 
non-linear $\sigma$-model with $\rm \theta=\pi$. For the coupling 
constant g one finds $\rm g=4\;(1-4J_{2}/J_{1})^{-1/2}$ \cite{houches}. 
Therefore the spinon velocity can be written as:
\begin{equation}\label{v}\rm
v=\frac{\pi {\rm J_{1}}}{2}\sqrt{1-\frac{4 {\rm J_{2}}}{{\rm J_{1}}}} .
\end{equation}
Logarithmic corrections, on the other hand, depend on a 
mass scale $\rm\Lambda$, which is generated dynamically in the $\sigma$-model 
, and for small g, ${\rm\Lambda}=1/a\; \rm \exp(-2\pi/g)$, 
where $a$ is the lattice spacing \cite{german}. 
Since $\rm T_{0}\propto\Lambda$,
\begin{equation}\label{T0}\rm 
T_{0}\propto\mbox{exp}\left[-\frac{\pi}{2}\sqrt{1-
\frac{4 {\rm J_{2}}}{{\rm J_{1}}}}\right]. 
\end{equation}
The results for v and $\rm T_{0}$ (Eq.(\ref{v}) and Eq.(\ref{T0})) are 
plotted in Fig.\ref{Fig_sigma} together with those obtained by finite size 
scaling.

In Table \ref{table1}, the estimated values v, $\rm T_{0}$ 
(following the above described procedure) are given for various
3-leg ladders. The spinon velocity increases, while $\rm T_{0}$ 
decreases with increasing ${\rm J_{\parallel}/J_{\perp}}$.

\begin{figure}[h]
\epsfxsize=80mm
\vspace*{-0.5cm}
\epsffile{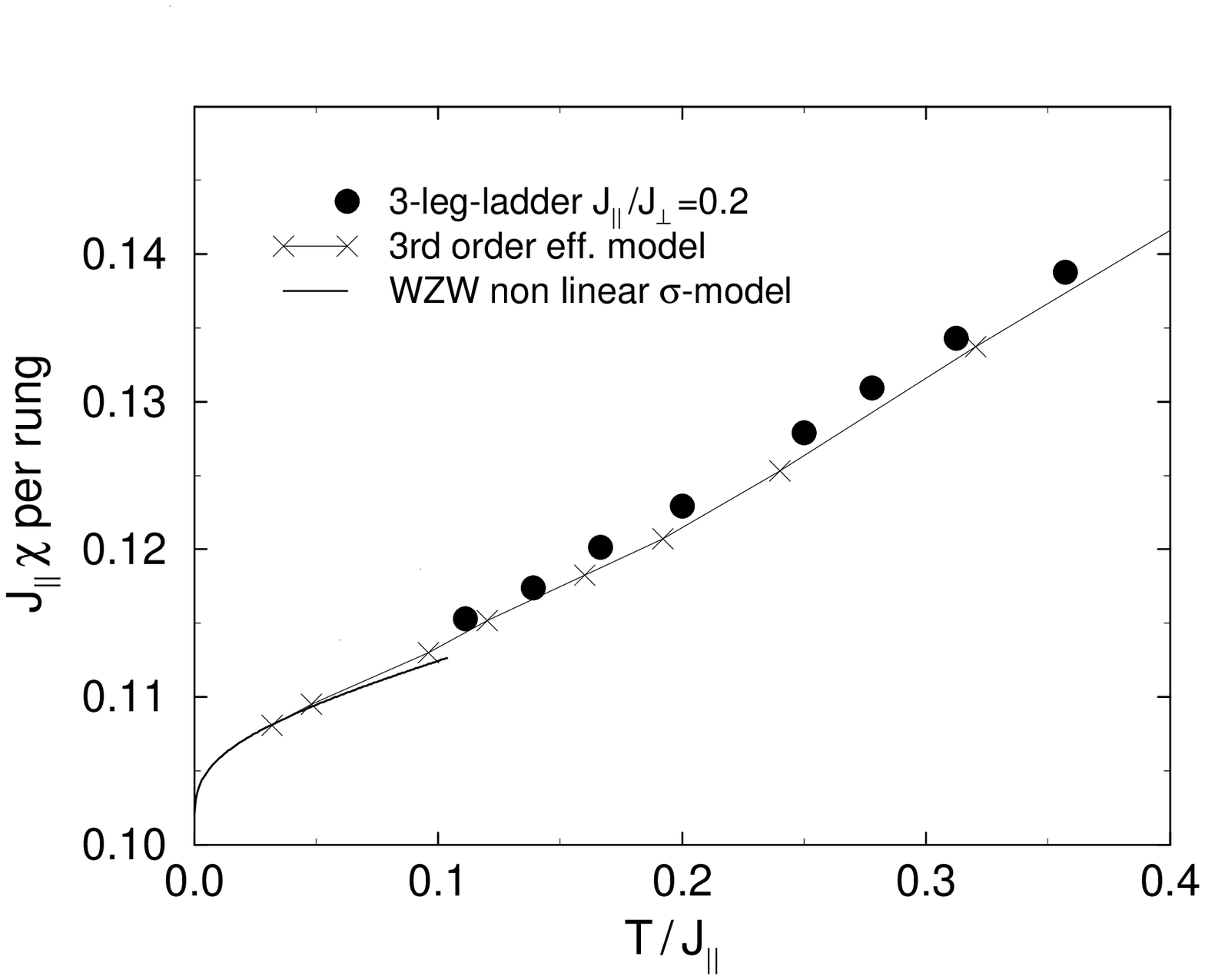}
\caption[*]{Susceptibility of the 3-leg ladder 
with $\rm J_{\parallel}/J_{\perp}=0.2$ and its effective models. 
The error bars are smaller or of order of the symbols.\label{Fig_susc02}}
\end{figure}

The values for v and $\rm T_{0}$ can now be put back 
into Eq.(\ref{1susc}) which gives the low temperature susceptibilities 
of the corresponding 3-leg ladders. This is shown at the example 
of ${\rm J_{\parallel}/J_{\perp}=0.2}$ in Fig.\ref{Fig_susc02}. 
Together with the QMC 
data for the 3-leg ladder itself and its third order effective model, we 
get the susceptibility for the 3-leg ladder with good precision on the 
whole temperature range from zero temperature to high T.

Greven et al. recently calculated the correlation 
length $\rm\xi$ of isotropic ladders \cite{greven}, using also the 
QMC loop algorithm. For $\rm T\ll T_{0}$ the 
inverse of the correlation length
in the WZW-model can be written as\cite{nomura2}
\begin{equation}\label{xsi}\rm
\frac{1}{\xi(T)}\approx T\left(2-\frac{1}{\ln(T_{0}/T)}+
\frac{1}{2}\frac{\ln(\ln(T_{0}/T)+1/2)}{\ln^{2}(T_{0}/T)}\right)
\end{equation}
Fitting the data \cite{greven} for $\rm n_{l}=3$ to the 
above form (Eq.(\ref{xsi})) also gives an estimate of $\rm T_{0}$. In the 
isotropic case we find $\rm T_{0}= 0.34J_{\parallel}$ which is in good 
agreement with our 
result (see Table \ref{table1}). For the 5-leg-ladder $\rm T_{0}$ is 
already $\lesssim\rm 0.1J_{\parallel}$.

Finally, we examine the static structure factor $\rm C(\pi,\pi)$, defined 
by ($\rm N_{s}$ = number of sites)
\begin{displaymath}\rm
C(k_{x},k_{y})=\frac{1}{N_{s}}\;\sum_{i,j,\tau,\tau'}
e^{i k_{x}(i-j)+ik_{y}(\tau-\tau ')}
<\vec{S}_{i,\tau}\vec{S}_{j,\tau '}> 
\end{displaymath}
and show that it can be calculated from the effective model without 
introducing additional physical parameters. $\rm C(\pi,\pi)$ can be written as
$\rm \sum_{i,j}(-1)^{i+j} <\vec{S}_{i}^{st }\vec{S}_{j}^{st }>/3L$, where
$\rm \vec{S}_{i}^{st }=\vec{S}_{i,1}-\vec{S}_{i,2}+\vec{S}_{i,3}$ is the 
staggered spin of one rung. In the limit ${\rm J_{\parallel}=0}$ the 
correlations of the staggered rung-spins is simply related to those of the 
uniform rung-spins by:
\begin{equation}\label{lambda}\rm
<\vec{S}_{i}^{st }\vec{S}_{j}^{st }>_{i\neq j}= 
\lambda <\vec{S}_{i}^{tot}\vec{S}_{j}^{tot}>_{i\neq j},
\end{equation}
with $\rm\lambda=25/9$ for all temperatures, 
where only the states in $\cal M$ are 
relevant. For ${\rm J_{\parallel}} \neq 0$ the ``wave function'' 
of the spin-1/2 
degree of freedom at each rung is spread out over a certain number of 
unit cells. On one hand, this gives rise to the longer range 
interactions, described above, and on the other hand, it affects the 
correlations $\rm <\vec{S}_{i}^{st }\vec{S}_{j}^{st }>$ 
and $\rm <\vec{S}_{i}^{tot}\vec{S}_{j}^{tot}>$. For large distances 
$\rm |i-j|\gg 1$, 
however, $\rm <\vec{S}_{i}^{st }\vec{S}_{j}^{st }>$ and 
$\rm <\vec{S}_{i}^{tot}\vec{S}_{j}^{tot}>$ are still related by 
Eq.(\ref{lambda}) but with a renormalized value of $\rm \lambda$. 

\begin{figure}[h]
\epsfxsize=75mm
\vspace*{-0.5cm}
\epsffile{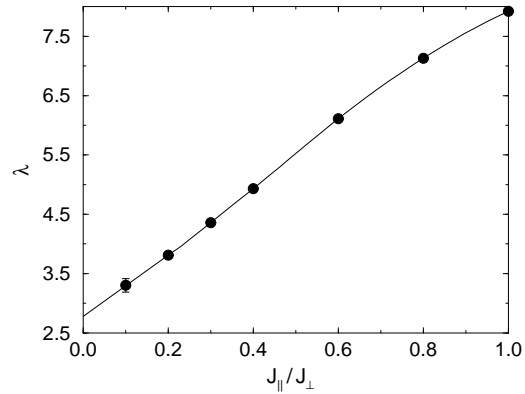}
\caption[*]{The fraction $\rm \lambda=<\vec{S}_{i}^{st }\vec{S}_{j}^{st }>
/<\vec{S}_{i}^{tot}\vec{S}_{j}^{tot}>$ at $\rm |i-j| \gg 1$ for 
different ${\rm J_{\parallel}/J_{\perp}}$. \label{Fig_lambda}}
\end{figure}

Here, we calculate $\rm \lambda$ numerically, using QMC simulations. The 
values are plotted in Fig.\ref{Fig_lambda} for 
different ${\rm J_{\parallel}/J_{\perp}}$ and extrapolating to 
${\rm J_{\parallel}=0}$ leads to a value $\rm \lambda=2.78$ which 
agrees with 
the analytical result $25/9$. At low T, the correlation 
length $\rm \xi\gg 1$ and $\rm C(\pi,\pi)$, as well as
$\rm C(\pi,0)=\sum_{i,j}(-1)^{i+j}<\vec{S}_{i}^{tot}\vec{S}_{j}^{tot}>/3L$
are dominated by correlations with $\rm |i-j|\gg 1$. Additionally, 
$\rm C(\pi,0)$ per rung of the 3-leg ladder differs only slightly 
from $\rm C(\pi)$ of the single chain. Consequently at low temperature, 
$\rm C(\pi,\pi)_{\rm 3-leg}=\lambda C(\pi,0)_{\rm 3-leg}\approx 
\lambda C(\pi)_{\rm single\ ch.}/3$. Greven et al. have
calculated $\rm C(\pi,\pi)_{\rm 3-leg}$ for ${\rm J_{\parallel}=J_{\perp}}$
and $\rm C(\pi)_{single\ ch.}$ \cite{greven}. From their data we determine 
$\rm C(\pi,\pi)_{3-leg}/C(\pi)_{single\ ch.}=2.55$, which 
is in good agreement with our value $\rm \lambda/3=2.64$ 
(see Fig.\ref{Fig_lambda}).

The above  considerations can be generalized to  an arbitrary 
odd-leg ladder with $\rm n_{l}$ legs. There are no 
qualitative differences.  The overall AF quasi-long range 
order, however, increases with increasing $\rm n_{l}$. Therefore, 
especially the ratio $|{\rm J_{2}/J_{1}}|$ of the effective 
Hamiltonian $\rm H_{eff}$ is larger. The logarithmic 
corrections increase markedly ($\rm T_{0}$ decreases) and 
the $\rm T \rightarrow 0$ behavior sets in at lower temperature as 
$\rm n_{l}$ increases. The zero temperature value $\rm \chi(0)$, on 
the other hand, is almost independent of $\rm n_{l}$. This implies 
that the spinon velocity of odd-leg ladders depends only slightly 
on the number of legs.

In conclusion, we have proposed an effective spin-1/2
chain model which
describes the low-energy properties of odd-leg ladders. 
The temperature dependence of $\rm \chi$, $\rm \xi$, and $\rm C(\pi,\pi)$
for the effective system is shown to be consistent with that of the 
original model at low enough temperatures. The effective model requires 
two parameters, e.g. a spinon velocity v and
an energy scale, $\rm T_0$. 
With an increasing number of legs, v does
not change, but $\rm T_0$ decreases
rapidly. The exchange interactions of the corresponding
effective model become longer ranged, and antiferromagnetic correlations
are enhanced.

The calculations were performed on the Intel Paragon of the ETH Z\"urich. 
The support of the Schweizerischer Nationalfond is gratefully acknowledged.

\vspace{-0.3cm}
\begin{thebibliography}{4}
\vspace{-1.2cm}
\bibitem{dagottoriera} E. Dagotto, J. Riera, and D.J. Scalapino,
 Phys. Rev. B \bf45\rm, 5744 (1992).

\bibitem{dagotto} For a review, see E. Dagotto and T.M. Rice,
 Science \bf271\rm, 618 (1996).

\bibitem{johnston} D.C. Johnston, J.W. Johnson, D.P. Goshorn,
 and A.J. Jacobsen, Phys. Rev. B \bf35\rm, 219 (1987).

\bibitem{hiroi} Z. Hiroi, M. Azuma, M. Takano, and Y. Bando,
 J. Solid State Chem \bf 95\rm, 230 (1991); M. Azuma et al.,
 Phys. Rev. Lett. \bf73\rm, 3463 (1994).

\bibitem{ich} B. Frischmuth, B.Ammon, M. Troyer, Phys. Rev. B (in press),
cond-mat/9601025.

\bibitem{evertz1} H.G. Evertz, G. Lana, and M. Marcu,
 Phys. Rev. Lett. \bf70\rm, 875 (1993).

\bibitem{evertz2} H.G. Evertz, M. Marcu, Nucl. Phys. B
 (Proc. Suppl.) \bf30\rm, 277 (1993).

\bibitem{wiese} U.J. Wiese, and H.P. Ying,
 Z. Phys. B \bf93\rm, 147 (1994).

\bibitem{greven} M. Greven, R.J. Birgeneau, and U.-J. Wiese, preprint, 
cond-mat/9605068.

\bibitem{eggert} S. Eggert, I. Affleck, and M. Takahashi,
 Phys. Rev. Lett. \bf73\rm, 332 (1994).

\bibitem{nomura} K. Nomura, Phys. Rev. B \bf48\rm, 16814 (1993).

\bibitem{houches} I. Affleck, Les Houches Lecture Notes, in: Field, Strings, 
and Critical Phenomena, ed. E. Brezin and J. Zinn-Justin (North Holland, 
Amsterdam, 1988).

\bibitem{german} G. Sierra, Jour. Math. Phys. \bf A29\rm, 3299 (1996).

\bibitem{nomura2} K. Nomura and M. Yamada, Phys. Rev. B \bf43\rm, 8217 (1991).


\end{thebibliography}
\end{document}